# On extracting sediment transport information from measurements of luminescence in river sediment

**Harrison J. Gray**[1,2]*, **Gregory E. Tucker**[2], **Shannon A. Mahan**[1], **Chris McGuire**[3], **Edward J. Rhodes**[3,4]

[1] *Cooperative Institute for Research in Environmental Sciences (CIRES) and Department of Geological Sciences, University of Colorado – Boulder, CO*

[2] *U.S. Geological Survey Luminescence Geochronology Laboratory, Denver Federal Center, Denver, CO*

[3] *Department of Earth, Planetary, and Space Sciences, University of California Los Angeles, Los Angeles, CA*

[4] *Department of Geography, University of Sheffield, Sheffield, S10 2TN, United Kingdom*

*corresponding author: harrison.gray@colorado.edu

**KEY POINTS**

- **We develop a model coupling sediment transport of fine sand and luminescence, in order to explain the patterns of luminescence observed in river sediment**
- **The model successfully reproduces the patterns of luminescence measurements in river systems**
- **Best-fit values from the model produce sediment transport information for fine sand within orders of magnitude from other river systems**

**ABSTRACT**

Accurately quantifying sediment transport rates in rivers remains an important goal for geomorphologists, hydraulic engineers, and environmental scientists. However, current techniques for measuring transport rates are laborious, and formulae to predict transport are notoriously inaccurate. Here, we attempt to estimate sediment transport rates using luminescence, a property of common sedimentary minerals that is used by the geoscience community for geochronology. This method is advantageous because of the ease of measurement on ubiquitous quartz and feldspar sand. We develop a model based on conservation of energy and sediment mass to explain the patterns of luminescence in river channel sediment from a first-principles perspective. We show that the model can accurately reproduce the luminescence observed in previously published field measurements from two rivers with very different sediment transport styles. The parameters from the model can then be used to estimate the time-averaged virtual velocity, characteristic transport lengthscales, storage timescales, and floodplain exchange rates of fine sand-sized sediment in a fluvial system. The values obtained from the luminescence method appear to fall within expected ranges based on published compilations. However, caution is warranted when applying the model as the complex nature of sediment transport can sometimes invalidate underlying simplifications.

**INTRODUCTION**

The rate of sediment transport by rivers is a key variable in understanding the evolution of landscapes [*Tucker and Hancock, 2010*], the behavior of rivers [*van Rijn, 1993*], the lifespan of reservoirs [*Syvitski et al., 2005; Papanicolaou et al., 2008*], and the impacts of development on sedimentation [*Syvitski, et al., 2005*]. Surprisingly, we have little ability to quantify sediment transport rates beyond hard-to-constrain analytical models and time-consuming tracer experiments [*Haschenburger and Church, 1998; Martin and Church, 2004; Bradley and Tucker, 2012* and references therein]. This knowledge gap reflects a lack of reliable field data with which to calibrate models, and uncertainties in the travel velocities and exchange rates of various grain sizes throughout a river system [*Papanicolaou et al., 2008*]. For this reason, it is important to explore possible connections between geomorphic process and material properties that may act as a proxy for these processes.

One such material property, luminescence, displays changes within river systems which may provide a means to obtain sediment transport information. Luminescence arises as a property of certain silicate minerals wherein bonding electrons excited by ionizing radiation become trapped in defects in a mineral's crystal lattice [*Rhodes, 2011*]. The trapped electrons occupy energy levels between the valence and conduction bands and remain stable until a source of energy such as heat or sunlight gives the electrons the energy needed to escape the trap, travel through the crystal (e.g., via the conduction band), and recombine with a radiative hole center, releasing photons in the process [*Rhodes, 2011*]. The emission of these photons due to energy from visible light is termed Optically Stimulated Luminescence (OSL) [*Huntley et al., 1985*]. A new method developed for potasssium-feldspar minerals uses infra-red light at a series of elevated temperatures and is termed post-Infrared Stimulated Luminescence (pIRIR) [*Thomsen et al., 2008; Buylaert et al., 2009*]. When pIRIR is measured at a series of elevated temperatures, different luminescence signals with different bleaching rates can be measured by a technique known as Multiple-Elevated-Tempurature post-infrared infrared stimulated luminescence (MET-pIRIR*; Li and Li, 2011*). The measurement of luminescence has been exploited as a geochronometer by the geoscience community, as common minerals such as quartz and feldspar can have trapped electrons removed by exposure to sunlight in a process known as "bleaching." This resetting by light exposure, and the subsequent buildup of luminescence due to background ionizing radiation when a mineral grain is buried, allows the determination of the elapsed time since last light exposure, which is taken to be equivalent to a depositional age [*Huntley et al., 1985*].

The downstream variation in luminescence of in-channel fluvial fine sand (90-250 µm grain size) has been documented in two studies. Stokes et al. (2001) observed that the equivalent dose of quartz OSL of sediment in the Loire River, France, displayed an overall decrease in luminescence with downstream distance from the river source. McGuire and Rhodes (2015a), using an MET-pIRIR protocol noted that the equivalent dose for various measurement temperatures also demonstrated a general decrease with downstream distance. Despite the striking difference in fluvial characteristics between the sites, the two studies revealed similar patterns: in both cases, luminescence tended to decrease downstream, at a rate that also decreased downstream. Furthermore, none of the samples collected showed complete bleaching, even though the sediment sampled was clearly subject to transport during high flows. The observed downstream decline of signal has been interpreted as a consequence of progressive bleaching during transport [*Jain et al., 2004; Gray and Mahan, 2015*]. While such an interpretation seems logical, it leaves several questions unanswered. What factors govern the rate of bleaching with respect to transport distance? Why does sand sampled from channel deposits retain a signal even when the material is clearly subject to contemporary transport? To what extent do variations in luminescence along a river reflect transport dynamics, such as the rate of channel-floodplain exchange or the virtual velocity of grains?

To begin to address these questions in a quantitative manner, we introduce a mathematical model which is similar to those for open channel flow and tracer transport [*Lauer and Parker, 2008a; Lauer and Willenbring, 2010; Pizzuto et al., 2014*], that describes the space-time evolution of quartz and feldspar luminescence signals in fluvial suspended sand. The model is then used to address three objectives. First, we compare model predictions with the data sets of Stokes et al. (*2001*) and McGuire and Rhodes (*2015*) in order to determine whether the model provides a consistent explanation for their observations. Second, we assess whether such a model, when fit to along-stream observations of luminescence, holds the potential to provide information about rates of and patterns of sediment transport---and if so, what additional constraints would be needed in order to maximize the value of such information. The third aim is to determine whether preliminary estimates of virtual velocity and sediment exchange rate derived from the two published datasets are broadly consistent with measurements from comparable fluvial systems. Collectively, these aims are intended to provide the first ingredients for a mechanistic theory of luminescence signal evolution in fluvial sand, and a first assessment of the potential use of such a theory for extracting information about sediment transport.

**MODEL FOR LUMINESCENCE IN SUSPENDED SAND**

Consider a channel control volume of width $w$, depth $h$, and stream-wise length $\Delta x$ (Figure 1). The total energy stored by the trapped electron concentration of suspended sediment within the control volume, $N_e$, is described by a basic conservation equation:

$$\frac{\partial N_e}{\partial t} = Q_{\text{upstream}} - Q_{\text{downstream}} - Q_{\text{bleaching}} - Q_{\text{deposition}} + Q_{\text{entrainment}} \qquad (1)$$

where the rate of change of total energy of trapped electrons $N_e$, (J) equals the sum of five energy fluxes of sediment (each with dimensions of Joules per time). These include influx by suspended-sand transport from upstream ($Q_{upstream}$), outflux by transport downstream ($Q_{downstream}$), loss of trapped electrons by sunlight bleaching ($Q_{bleaching}$), influx from entrainment of bed and bank sediment ($Q_{entrainment}$), and outflux by deposition ($Q_{deposition}$). The total energy within the control volume is:

$$N_e = \Delta x w h C \mathcal{L} \rho \qquad (2)$$

where $C$ is the sediment volumetric concentration, $\rho$ is the density of sediment, and $\mathcal{L}$ is the mean energy per kilogram expressed as sensitivity-corrected luminescence equivalent dose (J/kg). Equivalent dose refers the amount of absorbed radiative dose (J/kg) equivalent to produce the observed luminescence, and sensitivity-corrected means that the luminescence measurement is normalized by a small test dose of radiation such that luminescence measurements between different grains are comparable. The model is built around equivalent dose, rather than lab-measured luminescence intensity, as this controls for downstream changes in luminescence sensitivity [*Murray and Wintle, 2000; Pietsch et al., 2008*]. The model could also be built around sensitivity-corrected luminescence intensity; however luminescence intensity is measured in arbitrary units and we use equivalent dose instead because this quantity has defined units (J/kg) which helps demonstrates the statement of conservation of energy used in the model. Here we define the *mean equivalent dose* as the arithmetic mean of all aliquots for a sample. Because we are interested in the average bulk behavior among all grains, and no one grain of sand provides information about the transport histories of all grains, using a mean value allows us to average

the transport histories of many grains and obtain our desired estimates. The flux terms can be written as:

$$Q_{\text{upstream}} = whuC\mathcal{L}(x)\rho \qquad (3)$$

$$Q_{\text{downstream}} = whuC\mathcal{L}(x + \partial x)\rho \qquad (4)$$

$$Q_{\text{bleaching}} = \Delta x whC\mathcal{L}^*\rho \qquad (5)$$

$$Q_{\text{deposition}} = \Delta x wh f_D C\mathcal{L}\rho \qquad (6)$$

$$Q_{\text{entrainment}} = \Delta x wh f_E C\mathcal{L}_b\rho \qquad (7)$$

where $u$ is the 'drift velocity' [*Pizzuto et al., 2014*] of incoming sediment during transport, $\mathcal{L}^*$ is the rate of energy loss due to bleaching during transport (J/kg/s), $f_D$ is the fraction of the suspended sediment load that goes into storage per second (s$^{-1}$), and $f_E$ is the fraction of basal sediment entrained into the flow from storage (s$^{-1}$) with mean equivalent dose $\mathcal{L}_b$ (J/kg) (Figure 1). The variables, $f_D$ and $f_E$, represent deposited or entrained volumes of sediment normalized by the current volume of the suspended load. The volumetric flux of sediment $q_s$ (m$^3$/s) is defined as:

$$q_s = whuC \qquad (8)$$

Multiplying both sides by $u$, treating $u$ as constant in time and space, inserting equations 2 through 8 into equation 1, and dividing both sides by $\Delta x$ leads to the following equation:

$$\frac{\partial(q_s\mathcal{L})}{\partial t} = u\frac{q_s(x)\mathcal{L}(x)}{\Delta x} - u\frac{q_s(x+\partial x)\mathcal{L}(x+\partial x)}{\Delta x} - + q_s\mathcal{L}^* - f_D q_s\mathcal{L} + f_E q_s\mathcal{L}_b \qquad (9)$$

Taking the limit as $\Delta x$ approaches zero, applying the definition of the derivative and expanding all derivatives using the product rule leads to the complete equation:

$$q_s \frac{\partial \mathcal{L}}{\partial t} + \mathcal{L} \frac{\partial q_s}{\partial t} = -u\left[\mathcal{L}\frac{\partial q_s}{\partial x} + q_s \frac{\partial \mathcal{L}}{\partial x}\right] + q_s[f_E \mathcal{L}_b - f_D \mathcal{L} - \mathcal{L}^*] \quad (10)$$

Equation 10 describes conservation of energy stored as trapped electrons in the system with variable sediment transport rate ($q_s$) and sediment transport parameters ($u$, $f_E$, $f_D$). If Equation 10 were applied to a stream reach with a steady and uniform suspended-sediment load, the $\partial q_s/\partial x$ and $\partial q_s/\partial t$ derivatives would equal zero and $q_s$ would cancel from all terms. If we further consider a channel reach in which deposition and entrainment rates are approximately in balance, then $f_E = f_D = \eta$, where $\eta$ represents the sediment exchange rate (fraction of suspended sediment flux exchanged with storage centers per time). Under these conditions, the governing equation simplifies to:

$$\frac{\partial \mathcal{L}}{\partial t} = -u\frac{\partial \mathcal{L}}{\partial x} - \mathcal{L}^* + \mathcal{L}^\circ \quad (11)$$

$$\mathcal{L}^\circ = \eta(\mathcal{L}_b - \mathcal{L}) \quad (12)$$

Equation 11 is a kinematic wave equation with source/sink terms that are controlled by the sediment exchange ($\mathcal{L}^\circ$) and bleaching efficiency ($\mathcal{L}^*$). The parameter $\mathcal{L}^*$ (J/kg/s) represents the effective bleaching rate of luminescence during transport. Its value depends on how fast a luminescence signal is removed, which depends on the duration and intensity of sunlight, modulated by latitude, time of day and year, atmospheric conditions (cloudiness), and river conditions (flow depth, water turbidity). The parameter $\mathcal{L}°$ (J/kg/s) describes the effective flux of luminescence-bearing sediment into and out of active transport due to river erosion and deposition along the bed and banks. Luminescence measurements are typically made in the 90-250 μm grain size range, and therefore this model is applicable to the transport of fine sand. We elaborate on these assumptions in the discussion section.

**Definition of $\mathcal{L}^*$**

To solve for the virtual velocity ($u$) and erosion/deposition flux ($f_E$, $f_D$, $\eta$) parameters, it is necessary to constrain the loss rate of trapped-electron concentration due to sunlight exposure, $\mathcal{L}^*$. There are two conditions under which grains may be partially bleached: exposure to sunlight during subaqueous fluvial transport, and illumination of a thin layer of surface during periods between high flows when the drop in the water level exposes sediment on the higher parts of bars and banks. We define $\mathcal{L}^*_{fluvial}$ as the rate of equivalent dose decrease due to bleaching during subaqueous fluvial transport ($\mathcal{L}^*_{fluvial}$), and $\mathcal{L}^*_{surface}$ as the time-averaged rate of bleaching of a thin layer of deposited surface grains during low flows. The total rate of bleaching is then:

$$\mathcal{L}^* = \mathcal{L}^*_{fluvial} + \mathcal{L}^*_{surface} \qquad (13)$$

In this study, we assume that $\mathcal{L}^*_{fluvial}$ is significantly greater than $\mathcal{L}^*_{surface}$, because the latter involves only a small number of grains, and therefore include only on the former term. This assumption may or may not be applicable to all river systems [e.g. Porat et al., 2001] as some of the sediment entering the channel through entrainment might reasonably be expected to be material deposited during a recent event and exposed to sunlight during low-flow conditions. We explore the consequences of this assumption further in the discussion.

We take $\mathcal{L}^*_{fluvial}$ to be the derivative of equivalent dose with respect to time during sunlight exposure under fluvial conditions $\left(\frac{\partial D_E}{\partial t}\right)$. This derivative can be determined empirically from experiments in which aliquots of known dose are exposed to sunlight at various intervals, as described for example by the bleaching experiments of McGuire and Rhodes [*2015a*]. These data should be fit to an equation that can be differentiated to obtain $\left(\frac{\partial D_E}{\partial t}\right)$. We propose that the loss of equivalent dose due to bleaching could be described by a simple power-law equation such as:

$$D_E(t) = ((\beta - 1)k_t t + D_0^{1-\beta})^{\frac{1}{1-\beta}} \qquad (14)$$

$$\frac{\partial D_E}{\partial t} = -k_t D_E^{\beta} \qquad (15)$$

where $D_E$ is the equivalent dose (J/kg), $\beta$ is a non-dimensional constant, $k_t$ is an effective loss rate for equivalent dose (s$^{-1}$), and $D_0$ is the initial equivalent dose (J/kg). Equations 14 and 15 offer flexibility in fitting the data from these bleaching experiments. If possible, it is best to perform these experiments under light conditions expected during floods, such as under turbid water. Some possibilities include laboratory experiment [*Ditlefsen, 1992*], flume study [*Gemmell, 1985*], or experimentation in a turbid field environment [*Sanderson et al., 2007*]. Equations 14 and 15 are based on the assumption that the bulk bleaching rate of suspended grains in a well-mixed turbulent flow field has a similar power-law function as direct sunlight bleaching, but with significantly lower bleaching rate parameters ($k_t$ and $\beta$) than the direct-sunlight case. We show in the next paragraphs how we theoretically attenuate these bleaching rate parameters using simple subaqueous light attenuation physics.

For the Mojave River dataset, we use the bleaching experiment data of McGuire and Rhodes [*2015a*] to estimate $\mathcal{L}^*_{fluvial}$. Because they reported bleaching in terms of changes in luminescence intensity rather than equivalent dose, we need a method to translate between the two quantities. We use a saturating exponential of the form $y = A(1 - e^{-Bx})$, where A and B are constants, to express the relation between sensitivity-corrected integrated luminescence intensity *I* (arbitrary units) and equivalent dose (J/kg). Using this approach, equivalent dose, $D_E$, can be derived from integrated luminescence intensity, *I*, as:

$$D_E = -D_* \ln\left(1 - \frac{I}{I_{sat}}\right) \tag{16}$$

$$\frac{\partial D_E}{\partial t} = \frac{\partial I}{\partial t}\left(\frac{D_*}{I_{sat} - I}\right) \tag{17}$$

where $D_*$ is a growth parameter (J/kg) and $I_{Sat}$ is the integrated luminescence intensity at saturation (arbitrary units). Note that sometimes the observed response of luminescence as a function of dose is better described with functions other than an exponential, such as a combination of linear and exponential sometimes used in dating applications. Potentially another function could be used, but we do not explore this here for simplicity and the exponential function adequately describes luminescence groth for our purposes.. We use a saturating exponential for simplicity and to capture the saturating nature of luminescence as a result of dose. This approach is also used for the Loire dataset.

To define *I(t)*, we follow McGuire and Rhodes [*2015a*] and fit a power-law style equation (Equation 14) to their data, describing the change of luminescence intensity as a function of sunlight exposure time:

$$I(t) = ((\beta - 1)ft + I_0^{1-\beta})^{\frac{1}{1-\beta}} \tag{18}$$

where $\beta$ represents the non-dimensional decay order of the system, $I_o$ is the initial integrated luminescence intensity and *f* describes the loss rate for integrated luminescence intensity (s$^{-1}$). Note that Equation 18 describes the integrated luminescence photon counts versus sunlight exposure time for small aliquots.. Equation 18 adequately fits their experimental data ($R^2 = 0.95$). To account for the effects of river turbidity, we consider that the loss rate *f* scales directly with light intensity integrated over the sunlight spectrum:

$$f = \int_{\lambda_1}^{\lambda_2} \varphi(\lambda)\gamma(\lambda)\partial\lambda \tag{19}$$

where $\varphi(\lambda)$ is the incoming photon flux (sunlight) for a given wavelength $\lambda$ (photons/cm$^2$/nm) and $\gamma(\lambda)$ is the scaling of the loss rate *f* with photon flux under a given wavelength (cm$^2$/photons). Note that $\gamma(\lambda)$ is not strictly a photoionization cross-section but rather a value describing the change in loss rate of integrated luminescence intensity or equivalent dose due to variable sunlight intensity.

The penetration of sunlight through water, $\varphi(\lambda)$, can be described to a first approximation by the Beer-Lambert Law in photon-flux form for the attenuation of light in a fluid medium:

$$\varphi(\lambda) = \varphi_o(\lambda) e^{-\frac{z_{\text{eff}}}{z_*(\lambda)}} \tag{20}$$

The variable $z_*(\lambda)$ represents the attenuation of sunlight per wavelength in turbid water. The variable $z_{\text{eff}}$ represents the effective depth in the fluid at which a grain isolated at that depth would receive the same total amount of light as that expected for a grain undergoing random, turbulence-driven motion throughout the water column (see supplemental material for its derivation). We simplify Equation 19 and Equation 20 by assuming no wavelength dependence on the loss rate. With this simplification, $f = \gamma\varphi$ and the unattenuated loss rate $f_o = \gamma\varphi_o$ such that:

$$f = f_o e^{-\frac{z_{\text{eff}}}{z_*}} \tag{21}$$

which is then inserted into Equation 18. Equation 21 assumes that the magnitude of light intensity with depth exerts a stronger control on the loss rate than attenuation due to spectral filtering by absorption of some wavelengths of light by water. The filtering of higher energy wavelengths of light can lower bleaching rates [Sanderson, 2007; Reimann et al., 2015], but because we consider that fluid turbulence moves near grains toward the surface of the flow, which will not have significant spectral filtering, this is less important that the overall light intensity in water. We explore this assumption further in the discussion but note again that this assumption can be avoided by performing experiments to fit Equations 14 and 15.

**Definition of the basal sediment dose $\mathcal{L}_b$**

In order to implement the model, it is necessary to establish the luminescence input from entrainment of sediment, $\mathcal{L}_b$. We propose that this value can be determined either (A) empirically through measurements of sedimentary deposits near the river, (B) calculated from the sediment residence time distribution and background dose rate, or (C) calculated with a process-based sediment transport model. For this study, we follow the former approach (A) treat luminescence data from deposits near the river as $\mathcal{L}_b$ and use this value to calculate the characteristic storage timescale, $\tau_s$. To do this, the mean equivalent dose of sediment that appears to be on the verge of erosion is taken as $\mathcal{L}_b$ and divided by the background dose rate, $D_R$ (J/kg) to solve for $\tau_s$ (e.g. the relationship in Equation 22). Ideally, this empirical approach should involve a large number of samples to ensure accurate estimation of $\mathcal{L}_b$. This approach has the benefit of robustly determining $\mathcal{L}_b$ and can also examine its spatial distribution.

If the distribution of sediment residence time were known, the equivalent dose of basal sediment $\mathcal{L}_b$ could be taken as an expected value of sediment residence time $\tau_s$ times the background dose rate as a function of space and/or time $D_R(x, t)$:

$$\mathcal{L}_b = D_R(x,t) \cdot \tau_s \tag{22}$$

$$\tau_s = \int_0^\infty t_s p(t_s) dt_s \tag{23}$$

where $t_s$ is time spent in storage and $p(t_s)$ is the probability density function of sediment storage time. Note that this formulation assumes that the characteristic timescale of storage is below the saturation limit for the luminescence signal of interest. If the model is formulated in terms of luminescence intensity, then Equation 22 can be converted using equation 17. The expected value in Equation 23 depends on the probability density function $p(t_s)$ chosen to represent the system of interest. Determining $p(t_s)$ is beyond the scope of this paper, though significant research exists on this topic [e.g. *Bradley and Tucker, 2013* and references therein]. In the simplest case, p(t$_s$) could be assumed as an exponential distribution supported by field measurements. However, it is noted that this may not be appropriate for all systems. In some cases, the integral described in Equation 23 may not have a finite expected value if p(t$_s$) is governed by heavy-tailed probability distributions [*Bradley and Tucker, 2013*].

Finally, an alternative method to evaluate $\mathcal{L}_b$ would be to use a landscape evolution model constrained by field data. One example could be a meandering river system coupled with sediment transport modeling [*Bradley and Tucker, 2013*]. For illustrative purposes, we consider a simple system where a channel can access all of the storage center with equal probability. In this case, the rate of change in basal equivalent dose with time is

$$\frac{\partial \mathcal{L}_b(x)}{\partial t} = D_R + \eta f_V(\mathcal{L}(x) - \mathcal{L}_b(x)) \tag{24}$$

where $f_V$ is the ratio between the sediment volume in the channel and volume in the storage center. Combining Equation 24 with the transport Equations 11 and 12 produces a simple system where equivalent doses decrease during transport and increase during storage.

**Estimation of time-averaged virtual velocity *U***

In this section, we consider how time-averaged virtual velocity, defined as the velocity of sediment grains that alternate between periods of mobility and periods of storage [*Martin and Church, 2004*], relate to other parameters in the model. Virtual velocity may be quantified as:

$$U = \frac{total\ distance\ of\ travel}{total\ time\ in\ transport + total\ time\ in\ storage} \quad (25)$$

At any given moment, the vast majority of grains will normally reside in a storage center [*Meade, 2007*]. When storage time >> transport time, the ratio between the characteristic lengthscale sediment travels before deposition ($\ell_s$) and the characteristic timescale of sediment storage ($\tau_s$) provides an approximation of the time-averaged virtual velocity [*Martin and Church, 2004; Pizzuto et al., 2014*]:

$$U \cong \frac{\ell_s}{\tau_s} \quad (26)$$

Pizzuto et al., [2014] and Lauer and Parker [*2008b*] give relations for the characteristic transport lengthscale ($\ell_s$) which we modify slightly to produce the correct units from our model-derived values: :

$$\ell_s = \frac{u p_c q_s}{\eta\ p_f q_s} \quad (27)$$

where $p_c$ and $p_f$ are the relative concentrations of sand and silt grain sizes in the river channel sediment, here taken as equal for simplicity. The characteristic lengthscale is the downstream distance over which 1/e (~37%) of the suspended sediment volume enters storage [*Lauer and Willenbring 2010; Pizuto et al., 2014*]. Combining equations (26) and (27),

$$U = \frac{u/\eta}{\tau_s} \quad (28)$$

This relation indicates that if sediment drift velocity, exchange rate, and storage residence time were known, then one could also obtain virtual velocity. As a final note, the sediment exchange rate $\eta$ is converted from units of (s$^{-1}$) to (m$^{-1}$) or (km$^{-1}$) by dividing $\eta$ by the transport velocity $u$ and converting to meters or kilometers for direct comparison to the characteristic transport distance commonly used in sediment-budget studies [e.g. *Pizzuto et al., 2014*].

**Model behavior and predictions**

The model presented above demonstrates a series of behaviors and makes several predictions about the magnitude and spatial pattern of equivalent dose in river channel fine sand (Figure 2). Consider the case of a channel reach with uniform discharge, in which suspended sediment enters at the upstream end at a steady rate and with a constant initial luminescence signal. In this case, the model predicts that the equivalent dose in river channel sediment will tend to approach a steady value over some length scale; in other words, once suspended sediment has traveled beyond a certain distance downstream of the head of the reach, its mean equivalent dose becomes approximately uniform ($\partial L/\partial x \approx 0$). At this point, the suspended sediment has reached a state of equilibrium in which the influx of fine sand with high equivalent dose from storage centers is matched by the decrease due to sunlight bleaching. This is a theoretical state that a river would reach under constant forcing, that is, approximately constant sediment flux and approximately unchanging sediment transport parameters, $u$ and $\eta$, and if the sediment in the channel and storage center is well-mixed with regards to equivalent dose. A change in the bleaching rate ($\mathcal{L}^*$) or the equivalent dose of sediment eroded from storage ($\mathcal{L}_b$) will change the steady value. For example, a lower bleaching rate ($\mathcal{L}^*$), such as one might find by measuring a hard-to-bleach luminescence signal, implies a higher steady value than would a higher bleaching rate associated with easy-to-bleach signals (Figure 2). An increase in the background luminescence ($\mathcal{L}_b$), due for example to erosion of an older fill terrace, would be associated with an increase in the mean equivalent dose of in-channel sediment (Figure 2). Spatial perturbations to this steady value cause either an increase or decrease before returning to the steady value. For example, a tributary that introduces relatively unbleached sediment would cause a "spike" in the river channel mean equivalent dose, which eventually returns to the steady value further downstream (Figure 2). Similarly, entrainment of sediment with near-zero equivalent dose along a particular reach of the channel would cause a transient decrease in river channel mean equivalent dose for some distance downstream.

It is important to note that the model doesn't necessarily predict that luminescence starts at a high value and decreases downstream, but rather that the starting luminescence, whether bleached or unbleached, will increase or decrease until it reaches a steady value, which reflects a balance between loss of luminescence lost due to bleaching and influx of sediment from storage centers where regeneration can occur. The model solutions shown in Figure 2 treat the stored-sediment luminescence, $\mathcal{L}_b$, as a boundary condition. What happens when the luminescence is allowed to evolve dynamically, as described by Equation 24? To address this question, we couple Equation 24 with Equations 11 and 12 to produce a simple system wherein an initial influx of sediment with some equivalent dose enters from upstream and subsequently undergoes either transport, where it is bleached, or enters into storage where it is able to regenerate at some background dose rate (Figure 3). Both easy-to-bleach luminescence and harder-to-bleach luminescence (such as quartz OSL versus pIRIR$_{290}$) show similar patterns; the upstream reaches display either an increase or decrease in equivalent dose with transport distance until a steady-state condition is reached, downstream of which the equivalent dose is constant with distance. In this simple theoretical example, the mean equivalent dose in the storage center is higher than the mean equivalent dose in the channel for both signals. As in Figure 1, any changes due to additional sediment transport processes may change the steady-state value and introduce some downstream variation. However, the essence of the prediction, for both in-channel and floodplain sediment, is a gradual downstream decrease or increase in luminescence that asymptotes to a quasi-uniform value. For suspended sediment in the channel, this uniform value represents a balance between the addition of signal via entrainment of stored sediment, and the loss of signal to bleaching. For stored sediment, the predicted emergence of constant average luminescence downstream reflects a balance between signal acquisition from ionizing radiation, and signal loss due to dynamic sediment exchange with the channel.

**COMPARISON WITH FIELD DATA**

As a test of the model predictions, we compared the model with previously published measurements of luminescence/equivalent dose in river sediment for two very different river systems, the Mojave River in southern California, USA [*McGuire and Rhodes, 2015a*] (Figure 4), and the Loire in France [*Stokes et al., 2001*] (Figure 5). The Mojave River is a regional scale (~$10^2$ km long), single channel, desert ephemeral river which undergoes a large flood once every ~10 years [*McGuire and Rhodes, 2015b*]. In contrast, the Loire is a continental-scale (~$10^3$ km long), meandering temperate-climate perennial river that undergoes yearly flooding [*Stokes et al., 2001*]. Despite dissimilar hydrology, both datasets show an overall decrease in the measured luminescence with downstream distance (for feldspar MET-pIRIR$_{230}$ for the Mojave and quartz

OSL in the Loire), albeit with some notable deviations (Figures 4 and 5). In both datasets, the steady value at $\partial \mathcal{L}/\partial x \approx 0$ is greater than would be expected for a fully bleached sample. In the Mojave River dataset, the signals measured from the river sediment are much greater than what is observed from a fully sunlight-bleached sample [*McGuire and Rhodes, 2015a*]. In the Loire, channel equivalent doses at the downstream reaches demonstrate large variability but some data points are significantly lower than samples in the upstream reaches as discussed below.

In the Mojave River dataset, McGuire and Rhodes (*2015a*) collected samples from 0.3-0.5 meters depth in dry channel bar deposits with developed bedding structures. Equivalent dose for each sample was measured using the MET-pIRIR protocol (*Li and Li, 2011*) with post-IR temperatures of 95,140,185, and 180°C. The equivalent dose can be seen to follow a generally downstream-decreasing pattern with increasing transport distance (Figure 4). However, the downstream-most sample departs from this trend. Its location correlates with a downstream change in channel morphology from a relatively wide channel to a narrow reach with considerably higher and steeper valley walls [*McGuire and Rhodes, 2015a; 2015b*]. We interpret the data in Figure 4 as indicating that the upstream reaches of the Mojave follow a pattern of equivalent doses declining downstream toward a steady value whereas the downstream incised reaches demonstrate a potential increase in the equivalent dose of basal sediment $\mathcal{L}_b$ and/or represent a change in relative magnitude between the erosional exchange $f_E$ and depositional exchange $f_D$. Because the exact roles of either cause are not constrained, we treat the farthest downstream sample as an outlier and exclude it from the model. If information such as the $\mathcal{L}_b$ value in the downstream reaches were known, it would allow the modeling of this part of the system. However, the model presented here provides a consistent explanation for these observations (Figure 2).

For the Mojave River case, the loss rate $f$ and decay order $\beta$ (Equation 18) were directly measured through the bleaching experiments of McGuire and Rhodes [*2015a*]. We use the dose recovery data, in the form of parameters relating equivalent dose to sensitivity-corrected luminescence intensity, $D_*$, and $I_{sat}$, from McGuire and Rhodes [*2015a*] for Equations 16 and 17. The river during major transport events was taken to be turbid: $z_*$ was assumed to be 5 cm and constant across the sunlight spectrum. We estimated $z_*$ by observing video of the Mojave River during a major flood and estimating the depth to which submerged objects became obscured and converted this depth to a light attenuation constant using an empirical relation for desert lakes [*Idso and Gilbert, 1974*]. This is the best possible approximation considering the complexity of river turbidity [*Belmont et al., 2009*] and the fact that no theoretical relation exists for river turbidity [*Davies-Colley and Smith, 2001*] although empirical relations for select rivers exist [*Davies-Colley and Nagels, 2008; Julian et al., 2008*]. We assume that the in-channel sediment has undergone turbulent transport in these turbid conditions prior to deposition in channel bars from which our two datasets were sampled [*Stokes et al., 2001; McGuire and Rhodes, 2015a*]. The $\mathcal{L}_b$ value was determined from a terrace sample from McGuire and Rhodes [*2015a*] (Figure 4).

We applied the model to the Mojave River by running the model under the parameters described above. Only in-channel samples, including channel bars, were used to model the channel sediment luminescence. The terrace data were used to calculate the channel/storage center exchange ($\mathcal{L}°$). The model was run repeatedly and the parameters of *u* and *η* systematically changed on each iteration to find the best fitting run as determined by least-squares fitting (see supplemental material). After the best-fit values for *u* (transport velocity) and *η* (sediment exchange rate) were obtained, we used those values, and the terrace sample of McGuire and Rhodes [*2015a*] to obtain $\mathcal{L}_b$, and to calculate the characteristic transport lengthscale, $\ell_s$, characteristic storage timescale, $\tau_s$, and time-averaged virtual velocity, *U*. The model was run for each set of MET-pIRIR luminescence signals, which produced transport values that were internally consistent and within uncertainty. The values were then averaged and are reported in Table 1. Application of the model to the Mojave River results in a sediment exchange rate for fine sand of 17% ± 12% suspended load exchanged per kilometer, characteristic transport lengthscale of 6.9 ± 4.2 km, characteristic storage timescale of 3.6 ± 1.2 kyr and time-averaged virtual velocity *U* of 1.9 ± 1.4 m/yr (Table 1). These values are applicable to fine sand (90-250 µm).

The Loire dataset was obtained from Stokes et al., [*2001*] who collected samples by placing empty cans with a volume of ~333 cm$^3$ into unconsolidated sediment below 60 cm of water and immediately transferring the can to a light-proof polyethylene bag. Samples were collected at logarithmic spacing for the first 100 km and then at approximately 100 km increments further downstream [Stokes et al., 2001]. They used a single aliquot regeneration protocol with a single regeneration point of 4 Gray and a linear fit to the resulting growth curve. The measured equivalent dose shows a general decrease in equivalent dose with downstream distance similar to the Mojave River dataset (Figure 5). However, this pattern breaks in a location that roughly correlates with a shift from rural to urban land use as well as junctions with six large tributaries

[*Stokes et al., 2001*]. We do not apply the model to the full system of the Loire because of the assumptions used to derive the simplified model. These assumptions are approximately constant sediment load, constant bleaching rate, and approximately constant $\mathcal{L}_b$. Instead, we apply the model to the upper reaches of the Loire where we are more confident our assumptions are valid. For the case of the Loire, data to calculate $k_t$ and $\beta$ are not available. For illustration, we model the bleaching of luminescence intensity and convert to equivalent dose so that this dataset can be used to estimate sediment transport information. We estimate $f$ using the blackbody irradiation of the sun filtered through atmospheric and subaqueous conditions and the photoionization cross section of quartz OSL [*Singarayer and Bailey, 2004; Bailey et al., 2011*] and the decay order, $\beta$, is taken as 1. This is not strictly correct as $\gamma$ in Equation 18 is not the same as the photoionization cross-sections of quartz [i.e. *Jain et al., 2003*], and the measurement of quartz OSL in the Loire involved a combination of multiple quartz OSL components leading to a different $\beta$ [B*ailey et al., 1997*]. Further detail is provided in in the supplemental material. The e-folding length $z_*$ was taken to be 5 cm and constant across the sunlight spectrum for both rivers. We used the Modern and T1 terrace data from Colls et al. [*2001*] to estimate $\mathcal{L}_b$. Because no data on the dose-response curves for the Loire are available, we used data from the recent laboratory intercomparison quartz standard [*Murray et al., 2015*] to estimate $D_*$ and $I_{sat}$.

We applied the model to the Loire following the same order of operations as the application of the model to the Mojave River, using the parameters above to find the best-fit values of $u$ (transport velocity), and $\eta$ (sediment exchange rate). Application of the model results in a sediment exchange rate ($\eta$) for fine sand of 4.7 % ± 4.2% fraction suspended load exchanged per meter, characteristic transport lengthscale of 50 m, characteristic storage timescale of 1.2 ± 0.75 kyr and time-averaged virtual velocity, *U,* of 0.04 km/yr (Table 1). Transport velocity *u*, and sediment-exchange rate *η*, and associated uncertainties were determined by least-squares fitting (see supplemental material). These values are applicable to fine sand (90-250 µm). The large uncertainties in the Loire dataset lead to large uncertainties in the model fits. For the transport lengthscale, storage timescale, and virtual velocity, the relative uncertainty is over 100% and as such we show the resultant values simply for illustration. Note that in the application to both the Mojave and Loire, we are using single-aliquot data. This may be preferable to use aliquots consisting of a large number of grains in order to capture the histories of many grains to describe the bulk behavior of all grains.

## DISCUSSION

### Model application to the Mojave and Loire

We are very encouraged by these results. To our knowledge, this is the first mechanistic model of luminescence in suspended river channel sediment that provides a self-consistent explanation for the observed patterns of luminescence in river sediment. Within the model domain, it is able to reproduce the spatial distributions of luminescence in river sediment from a conservation of energy and mass first-principles approach. The best-fit parameters from application of this model are interestingly close to the results found from other systems and demonstrates the potential for deriving information about fine-sand sediment transport from luminescence measurements (Table 1).

The model's initial apparent successes suggest that further study is warrented. Evaluating the accuracy of the method is complicated by the lack of data on exchange rates, characteristic length and time scales, and virtual velocities for fine sand [*Parsons et al., 2015*]. Virtual velocity has largely been used to describe the movement of pebble and cobble tracers that travel as bedload [*Hassan et al., 1992; Haschenburger and Church, 1998; Bradley and Tucker, 2012*]. However, recent work also uses the virtual velocity concept to understand long-term suspended sediment transport [*Pizzuto et al., 2014; Parsons et al., 2015*], particularity because of the tendency for contaminants to sorb onto fine sediment [e.g *Pizzuto, 2014*]. Pizzuto et al. [*2014*] present the largest compilation to date of the exchange rates, characteristic length and time scales, and virtual velocities for fine sand resulting from many studies in the Mid-Atlantic region of the northeastern United States. Their range in values is compared to our results in Table 1. Our results approximately fall within the range of magnitudes obtained in their study with some exceptions. Our results from the Mojave River dataset are concordant with all value ranges except the storage timescale. The results from the upper Loire dataset are outside the ranges and seem to present values representing a system with much slower sediment transport than seen in the Pizzuto et al., [*2014*] data. On the other hand, Pizzuto et al. [*2014*] note that the expected true range of these values across all rivers may span orders of magnitudes beyond what they observe. The concordance of the Mojave River results with the previously published data is encouraging, particularly because of the inclusion of bleaching experiment data, data which were not available for the Loire dataset. A more rigorous evaluation of the luminescence-derived sediment transport information would require collection of luminescence data and independently derived virtual-velocity estimates for the same fluvial systems. Such a paired study would also make it possible to explore the role of parameters such as hydraulic geometry, basin erosion rate, and climate.

The applicability of the simplified model (Equations 11 and 12) depends on a series of assumptions. In order to make predictions on the general trend of the luminescence in channel sediment, we have to make simplifying assumptions. Incorporating every single process in a river system is an impossible task and not the point of geomorphic modeling, which seeks to identify process signals from background noise. Instead, the goal is to obtain the minimum level of model complexity needed to produce robust predictions on the process of interest. These assumptions are: (1) that the majority of bleaching of trapped charge occurs during fluvial transport; (2) that characteristic transport and storage lengthscales and timescales for a river system have finite averages and/or variance; (3) that steady-state approximations are appropriate over suspended sediment transport timescales; and (4) that no significant geomorphic disequilibria such as major changes in sediment supply are occurring over the timescales of fine sediment transport. These assumptions are valid under certain conditions that must be upheld or the model modified to accommodate these changes.

**Bleaching during transport**

Two central assumptions on the bleaching of luminescence were used in this model. First, we assume that removal of trapped charge largely occurs during in-channel transport by water. The counterpoint to this assumption is the possibility that the majority of bleaching occurs while sand is exposed at the surface of depositional units [*Porat et al., 2001; Gray and Mahan, 2015*]. The relative role of surface bleaching versus transport depends on the size of the river system, the magnitude/frequency of transport, the turbidity of the water, and the effective depth to which sunlight can penetrate stationary sediment.

The relative volumes of surface-bleached material versus in-transport bleaching depend on the relative scaling of bleaching depth versus scour depth. Investigating luminescence as a surface exposure chronometer, Sohbati et al. [*2012*] found that the depth to which sunlight penetrates and bleaches Navajo Sandstone is on the order of 2-4 mm for ~80 years of sunlight exposure and 4-8 mm for 713 ±61 years [*Sohbati et al., 2012*]. For granite, Sohbati et al. [*2012*] modeled bleaching at approximately 12.5-17.5 mm depth for $10^2$ years and approximately 15-20 mm depth for $10^3$ years. Whether the bleaching depths of unconsolidated sediment follow this pattern is not yet known but may be on the same order of magnitude. Surface exposed sediment may also be bioturbated to deeper depths although the rate and magnitude of this depends on local biota. We estimate that bleaching in sediment reaches depths of 1-20 mm for the limited time sediment is in temporary in-channel storage such as bars. The expected depth to which sediment will be scoured and mobilized is a fractional power function of discharge, which depends on grain size and local river geometry [*Leopold et al., 1966; Hassan et al., 1998; Lu et al., 2012*]. Bleaching depths are likely small compared to typical scour depths [*Leopold et al., 1966*] and small or annual floods may largely mix surface bleached sediment with other surface bleached sediment. Scour depths for large floods in midsized rivers similar to those in this study can be on the order of 0.5 to 1-2 meters. If the depth expected to have been bleached by surface exposure is on the order of a few percent of the scour depth for characteristic "effective discharge" floods [*Wolman and Miller, 1960*], it could be assumed that the transport bleaching is dominant. This is further supported by the observation that smaller yearly floods move negligible volumes of sediment when compared to the larger characteristic floods that move the majority of sediment [*Nash, 1994*]. If significant, the effect of the surface bleaching could potentially cause an underestimation of the modeled transport velocity and an overestimation of the sediment exchange rate due to more efficient bleaching than expected. In this case, a value for $\mathcal{L}^*_{surface}$ is needed to reflect entrainment of this bleached sediment. However, the best method to obtain this value is not immediately clear.

The second assumption used in this model is that while fine sand is in suspended transport, the role of spectral attenuation by water is less important than the magnitude of sunlight intensity. Evidence for a dependence of bleaching rate on wavelength has been previously explored and shown to have a notable effect [*Singarayer and Bailey, 2004; Sanderson et al., 2007; Kars et al., 2014*]. However, Ditlefsen [*1992*] noted that clear water played little role in the bleaching of potassium-feldspar OSL and thermoluminescence compared to sunlamp exposure. The role of wavelength dependence in this model is complicated by the rapid turbulent mixing of fine sand in suspension. The grain sizes commonly used in luminescence dating and the ones used in this study (90-250 µm) travel as suspended load even in low flow due to their low Rouse numbers [*Rouse, 1937*] and low settling velocities [*Ferguson and Church, 2004*]. The Rouse number, *R*, is a non-dimensional number that expresses the ratio of gravitational settling to turbulent upward momentum:

$$R = \frac{\omega_f}{ku_*} \tag{29}$$

where $\omega_f$ is the particle settling velocity, $u_*$ is the shear velocity, and $\kappa$ is von Karmen's constant (0.41). For R ≥ 2.5, sediment is dominantly bed-load material, for R = 2.5-1.2, sediment is in partial suspension (saltation); for R =1.2-0.8, sediment is in full suspension; for R >0.8, sediment travels as wash load with minimal chance of bed contact [*Rouse, 1937*]. Here, we interpret suspended grains as those that take an extended distance of transport in between moments of contact with the bed. For a flow one meter deep ($h = 1$), with a channel slope $S = 0.001$, a shear velocity calculated from the depth-slope product ($u_* = \sqrt{ghS}$), and settling velocities calculated from Ferguson and Church [*2004*], the coarsest grain size used in this study (250 µm) will have a Rouse number of approximately 0.8: well within the suspended range. As the depth of the flow or slope increases, or the grain size decreases, the Rouse number will only decrease as the flow is able to produce stronger turbulence. Furthermore, if we consider that the majority of sediment is transported in large floods which have significantly greater discharge than the annual flood [*Wolman and Miller, 1960; Nash, 1990*], the grain sizes used in this study (90-250 µm) should be considered as suspended sediment. Turbulence will cause grains to move throughout the water column [*Argall et al., 2004; Man and Tsai, 2007*] with brief, but potentially frequent, exposure to light at the flow surface.

Because grains rapidly move from the bed to the surface of a flow, their bleaching history integrates periods of high-intensity, low spectral attenuation near the surface, to low-intensity, high spectral attenuation near the bed. The effect of turbulence on the bleaching rate was also observed by Ditlefsen [*1992*] and Gemmell [*1985*] where both witnessed a lowering in bleaching rate with increased turbulence, suggesting that turbulence brings more sediment into the flow and increases the water opacity despite also elevating grains closer to the surface. We cannot account for this with the current bleaching experiment data of McGuire and Rhodes [*2015a*] because the experiment was not performed in turbid water. However, if we assume that the magnitude of light intensity is more important than the role of spectral attenuation on bleaching rate, we can use Equation 18 as a first-order approximation. Further research into the relative role of turbulence versus spectral filtering on bleaching rates would help to better evaluate this assumption. Finally, assume that turbidity, represented by $z_*$, is constant. We base this on the expectation that the higher flows, during which large amounts of sediment are transported, will typically be turbid. However, we acknowledge that significant complexity exists with respect to turbidity [*Belmont et al., 2009*] and its effect on luminescence bleaching [*Gemmell, 1997*] such that further study is warranted.

It is important to point out that the model provides a framework for understanding the general trend of the mean equivalent dose with increasing transport distance rather than the random fluctuations in equivalent dose potentially due to smaller-scale processes such as depositional mechanism across a point bar [*King et al., 2014a; Cunningham et al., 2015a*]. Although the model can be modified to include the effects of processes such as erosion of older terraces and tributary input with high stored-sediment luminescence, $\mathcal{L}_b$ (Figure 2), the potential effects of smaller scale processes must be considered. In an elucidative series of papers, King et al. [*2013; 2014a; 2014b*] found that for glaciofluvial braid-bar systems, the dispersion in luminescence measurements within a single bar could be greater than the change in luminescence downstream over a 1-10 km study reach. We suggest their data also imply that both the dispersion and magnitude of luminescence intensity decrease with distance, especially at reach-scale (10-100 km) transport distances, potentially consistent with the results of our model, although their sediment system was notably different from those considered here. Cunningham et al. [*2015a*] discovered a correlation with the proportion of bleached grains versus height above the low-flow water level for a South African bedrock river, which they interpreted as indicating deposition by large turbid floods, as opposed to clear water during low flows. However, significant scatter seems to be present when these variables are compared in the lower Rhine [*Cunningham et al., 2015b*]. Porat et al. [*2001*] found that for a flash-flood-driven ephemeral river in southern Israel, the variability in equivalent dose within individual deposits obscured any potential downstream trend in their 800 m study reach. However, their suggestion that an 800-m reach is too small to see these trends is consistent with the parameters obtained during our application of the model, which demonstrates that downstream trends are apparent att the scale of tens of kilometers.

It should be noted that the model is here applied to all grains in the 90 – 250 µm range under the assumption that the bleaching rates across grain sizes is similar and that transport information across these sizes can be averaged. Grains in this range have been observed to have size-dependent residual doses in modern sediment [*Olley et al., 1998*], with the interpretation that coarser grain sizes are generally better bleached than finer sizes [*Wallinga, 2002; Truelsen and Wallinga, 2003; Rittenour, 2008*]. This is counterintuitive as it would be expected that finer grain sizes undergo greater fluid suspension than coarse grains and should therefore be better bleached [*Wallinga, 2002; Rittenour, 2008*]. However, this observation may be explained by the finding that finer grain sizes tend to have higher exchange rates and shorter transport length scales than coarse grain sizes [*Lauer and Willenbring, 2010*]. This would mean that finer grain sizes have greater probability of being deposited in floodplains for longer periods and regenerating signal during deposition. The relative difference in sunlight exposure due to greater suspension of finer grains may actually be insignificant because all grains in the 90-250 range will have low Rouse numbers and greater time spent in fluid suspension during the large floods that move the majority of sediment [*Wolman and Miller, 1960*]. Another possibility is that the greater residual doses seen in finer grain sizes reflect differences in intrinsic bleachability of a luminescence signal at that grain size. McGuire and Rhodes [*2015a*] noted that bleaching experiments across the 125-250 µm grain size range seem to show consistent behavior such that characterizing the bulk bleaching behavior of grains in this range is sufficient for our purposes. Flume experimentation may be necessary to conclusively test whether potential processes such as clay flocculation [e.g. *Lepper, 1995*] lead to differential grain size bleaching rates in turbulent flow. However, bleaching rates do not currently seem to be a major source of uncertainty in our model results.

Finally, in order to use a luminescence signal for this method, the luminescence intensity or equivalent dose must decline in an approximate and consistent manner with progressive sunlight exposure. Signals such as OSL, IRSL, post-IR IRSL, MET-pIRIR, and TL seem to follow this pattern sufficiently during sunlight bleaching experiments (*Reimann et al., 2014; McGuire and Rhodes, 2015a; Colarossi et al., 2015*). The choice of luminescence signal may depend on the environment and the scale of interest. Fast to bleach signals, such as OSL, may bleach so rapidly that the mean equivalent dose is so close to zero that large uncertainties result in the derived sediment transport values. Slow to bleach signals, such as TL or high-temperature post-IR IRSL, may have trouble reaching steady-state dose conditions (i.e. $\partial L/\partial x \approx 0$) due to changes in input sediment ($\mathcal{L}_b$) because of the longer distances needed to bleach sediment than easier to bleach signals (Figure 2). However, it is worth noting that harder to bleach signals may produce more consistent long distance patterns as the faster bleaching rate of signals such as OSL can lead to large statistical dispersion under variable light exposure and greater inter-sample noise. Further research and experimentation will be needed to access which signal is appropriate for which environment. However, this may be an advantage as different MET-pIRIR signals could potentially allow one to 'fine tune' for the environment of interest.

**Sediment residence time and characteristic scales**

A key advantage of this method is that both the characteristic lengthscale of transport, $\ell_s$, and the characteristic timescale of storage, $\tau_s$, can be estimated from the luminescence in channel sediment and nearby deposits. One of the central assumptions in the model is that both $\ell_s$ and $\tau_s$ of sediment transport can be captured by definable averages for the majority of grains. The model implements this assumption in two ways. The first is the assumption that the characteristic transport lengthscale, $\ell_s$, accurately defines the distances average grains travel between deposition in long-term storage centers. The second assumption is that the time each grain spends in storage, the characteristic storage timescale, $\tau_s$, can also be captured by a definable average. The validity of these assumptions rests on the probability distributions controlling sediment transport and sediment storage [*Furbish et al., 2012*].

Sediment in transport is commonly thought of as an ensemble of particles that undergo periods of motion and periods of rest [*Furbish et al., 2012*]. Probability distributions can be used to describe the stochastic nature of these sediment transport episodes, which together can be used to define the nature and virtual velocity of transport [*Haschenburger and Church, 2001; Bradley and Tucker, 2010; Furbish et al., 2012*]. This concept can be found in studies of bedload transport [*Furbish et al., 2012; Rosenberry et al., 2012*] and saltating particles in wind [*Anderson*

*and Haff, 1988; Valance et al., 2015*]. However, an understanding of the transport episodes of suspended sediment transport is surprisingly lacking [*Parsons et al., 2015*]. The probability distribution of suspended sediment transport distances is uncertain, so we can only hypothesize. It is debatable what the effects of differing suspended sediment transport distance probability distributions would be on the mean equivalent dose of channel fine sand. Grains of sand in rivers should have greater cumulative sunlight exposure with cumulative distance given that fluid turbulence should move the grain into the photic zone near the water surface repeatedly. This scaling of sunlight exposure with transport distance would mean that the decrease in luminescence for a grain should also scale with transport distance. However, this scaling breaks down if the grain becomes fully bleached. If the suspended sediment transport distance distribution favors long transport distances, grains will be advected without recording further transport distance. However, if the mean of the distance of transport distribution is short, then the scaling holds and this may support the characteristic transport lengthscale method used in the model. Whether a mean of transport distances can be captured by a luminescence signal depends on the signal's bleaching rate with fast to bleach signals, such as IR50 and quartz OSL, having a higher probability of being completely removed before a transport episode of a grain ends. For this reason, slower to bleach luminescence signals are advantageous as they are more likely to capture the mean transport distances than faster-to-bleach signals. Note that in this framework, we consider only episodes of transport between residence time in long-term storage where significant signal regeneration can occur. This restricts our consideration to the transport distances during characteristic floods. Short rests at the base of the channel may not matter unless the water is sufficiently clear to allow bleaching at the bed. Further study of the influence of suspended sediment transport distance probability distributions on luminescence could be approached with particle-based modeling and single-grain measurements.

Our second assumption, that the characteristic storage timescale, $\tau_s$, can be captured by a definable average, depends on the probability distribution of the long-term "rest" or storage times of fine sand in natural river systems. We assume that there exists some characteristic mean timescale of fine sand storage such that on average, a particle rests for this amount of time before the next episode of transport. This assumption is also used in many estimates of time-averaged sediment virtual velocity [*Pizzuto et al., 2014* and references therein] and is implicitly made in many studies that collect geochronologic samples from fluvial landforms and assume the resultant age is representative of the landform. In counterpoint, the possibility exists that the distribution of fine-sand residence times are such that recently deposited fine sand has a greater probability of re-entrainment than fine sand that was deposited earlier, leading to a power-law ("gambler's ruin") sediment residence time distributions [*Tsai et al., 2014*]. Another possible effect is an 'erosion hazard' function in which re-entrainment is most likely at early timescales and at the longest timescales as channel migration moves back and forth between valley walls [*Bradley and Tucker, 2013*].

The effect of residence time distribution on the equivalent dose of channel fine sand would be expressed through modification of the basal sediment dose $\mathcal{L}_b$. If a channel reoccupies a location where it has recently deposited material, the $\mathcal{L}_b$ value will be lower due to smaller luminescence regeneration (say, tens to hundreds of years) than if the channel occupies a location that it had not occupied for a long time (say, hundreds to thousands of years) and significant luminescence regeneration occurred. The effect of floodplain residence-time distribution on the luminescence characteristics of channel sand deserves further study. The breadth and nature of this residence-time distribution should depend, among other things, on the sediment exchange rate, valley width, and channel migration rate [*Bradley and Tucker, 2013*]. Higher sediment exchange rates and/or meandering rates will more likely allow the channel to reoccupy a previous location [*Lancaster and Casebeer, 2007; Lauer and Parker, 2008b; Wickert et al., 2013*] and potentially move from one side of the valley back to the other. An additional complicating factor is that these rates may change with downstream distance [*Constantine et al., 2014*], which would alter the probability distribution of storage times. Furthermore, an anastomosing system, such as a braided channel, may have different residence times and exchange rates for each channel braid, leading to further complications. As mentioned previously, particle-based numerical process modeling may provide a way to assess the role of the sediment residence-time distribution. Furthermore, using a coupled landscape evolution and sediment transport model could provide insight into how processes such as river meander into older deposits may affect the pattern of channel equivalent dose.

**Steady-state approximations and geomorphic disequilibrium**

To apply the model for the field cases, we assume that the fine sand flux in the river channel, $q_s$, is constant in space and time over the study areas. This assumption allows for a direct application of the model without requiring additional information on the fine sand flux in the river in order to constrain transport velocity, $u$, and exchange rate, $\eta$. The assumption is, in effect, a statement that there are no significant spatial or temporal variations in the state of sediment supply of fine sand throughout a river system over the timescales relevant for fine sand transport from original erosional source to final depositional sink. These timescales are difficult to constrain, however, as storage times dominate the time a grain takes to cross from source to sink [*Pizzuto et al., 2014*] and these storage times can span the range of hundreds to hundreds of thousands of years [*Lauer and Willenbring, 2010*]. We propose that the timescale of interest, $\tau_{system}$, for this model is the expected time for a grain to travel the length of the river system, which is equal to the length of the river system, $\ell_{system}$, divided by the virtual velocity, *U*:

$$\tau_{system} = \frac{\ell_{system}}{U} \tag{30}$$

Equation 30 shows how a longer river, or a lower virtual velocity, provides a larger timescale of integration. For the Mojave River study area, our estimated virtual velocity implies a timescale on the order of $10^4$ years; for the Loire study area, the timescale is on the order of $10^5$ years. As the timescale and lengthscales of the observed system increase, the assumption of a uniform sediment load becomes more tenuous. Nonetheless, our luminescence-derived virtual velocity estimates imply that the characteristic timescales for fine-sand transit in these two river systems integrate over a range of climate conditions, which is consistent with the findings of sediment-budget studies in other rivers [*Pizzuto et al., 2014* and references therein]. This indicates that the results presented here represent fine sand transport information (*u, η*, and derived values) averaged over a long period of time and potentially over periods of geomorphic transience. How this information is averaged by transport processes [*Willenbring et al., 2013*] and whether changes in climate could be measured on this system [*Jeromack and Paola, 2010; Willenbring and von Blanckenburg, 2010*] remain significant research frontiers. However, we note that the assumptions involved in applying the model are not dissimilar to those used for analogous methods such as beryllium-derived catchment erosion rates, which form the basis for many successful studies [*Portenga and Bierman, 2010* and references therein], or sediment transport modeling [*Lauer and Parker, 2008a; Lauer and Willenbring, 2010; Viparelli et al., 2013; Belmont et al., 2014; Pizzuto et al., 2014*].

The presence of changing sediment loads, such as due to geomorphic landscape transience, does not necessarily invalidate the model. Rather, it requires that the additional information be taken into account in the theoretical framework. Observing a roughly constant value with downstream distance shows that a steady balance of bleaching and exchange is occurring and that these values can be quantified. Alternatively, if the luminescence versus downstream distance can be observed to be approximately increasing or decreasing, it may also be possible to quantify sediment transport information. Observations which show that large and frequent changes in the luminescence are occurring with downstream distance may indicate that the system is highly influenced by random influxes of unbleached/bleached sediment and modeling the patterns of luminescence and sediment transport difficult. Note that this observation is only relevant for samples taken in a consistent geomorphic location such as deep >30 cm sediment in channel bars. Sediment taken from banks may not be comparable with deep channel bar sediment or shallow channel sediment and so forth. The general 1D conservation law (equation 10) allows for the possibility that fine-sand sediment load, $q_s$, varies in time and space. Likewise, the model can accommodate changes in the dose of eroded fine sand, as might be expected for a channel that erodes material of different ages in different reaches (Figure 2). For example, the Mojave River dataset shows an increase in mean equivalent dose in the farthest downstream reaches of the channel (Figure 3E). This stretch of the river is characterized by an incised channel, and it receives water and sediment contributions from tributaries that have incised into alluvial fans. One might therefore expect an influx of material with a differing equivalent dose (new $\mathcal{L}_b$) into the main channel along this reach, which could explain the higher trapped charge concentration that we observe. If the $\mathcal{L}_b$ of the fine sand in the incised terraces were known, one could determine whether there is a difference in sediment-exchange rates within this incised reach as compared to the stable reaches upstream.

For future applications and research, we propose a series of approaches. First, we recommend that the parameter $\mathcal{L}^*$ be determined empirically through bleaching experiments in natural sunlight. As noted previously, the Mojave River dataset includes the bleaching experiments of McGuire and Rhodes [*2015a*], which allows for the surface value of *f* to be determined and does not require the extensive parameter estimation involved with the Loire data. The best-case scenario would involve exposing river sediment with a known equivalent dose to sunlight conditions typical of sediment in active transport. Subaqueous bleaching experiments made at the effective depth of transport (see supplemental material) would provide a useful value of *f* and would help limit the assumptions necessary to produce a velocity estimate. We also recommend large numbers of samples taken from river sediment, as this would greatly improve the estimates of sediment transport. In particular, sample collection that focused on observing long-distance trends in $\partial\mathcal{L}/\partial x$ would improve results. Additional information on the luminescence of the eroding material $\mathcal{L}_b$ and sediment residence time probability *p(ts)* would further improve estimations. Particle-based numerical modeling incorporating sediment suspension and/or channel migration mechanics could provide an avenue to explore these effects. The detection and measurement of very slow-to-bleach luminescence signals may improve the utility of the model for larger-scale estimates of sediment transport. As a final note, this model may also be applicable to fine silt (4-11 µm) which is also routinely measured in luminescence dating.

**CONCLUSIONS**

The model presented here demonstrates the potential of luminescence as a sediment transport indicator of fine sand. A simple model derived from conservation of energy (stored as luminescence) and conservation of mass (as sediment) can reproduce the downstream patterns of luminescence observed in two river systems, the Mojave River in southern California, USA and the Loire in southern France. Application of the model can produce sediment transport information, i.e. characteristic transport length and time scales, storage center exchange rates, and time-averaged virtual velocity values that appear to be realistic based on values observed in other river systems. The model requires a series of assumptions that may or may not be valid in all circumstances and must be considered thoroughly, and properly accommodated when possible.  However, deviations from the expected steady state conditions described by the model may help locate and interpret geomorphic disequilibrium. This study indicates that luminescence may hold significant utility towards obtaining sediment transport information from river systems and provide a potential new method to collect this data.

**FIGURES**

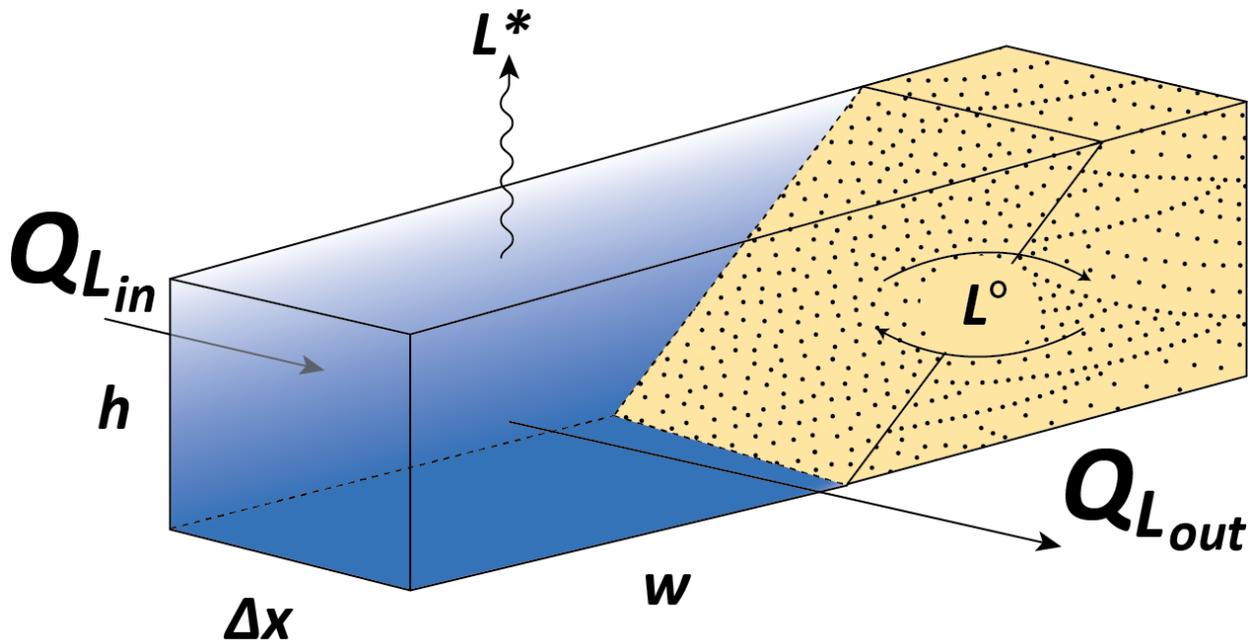

**Figure 1:** Definition diagram for the model used in this study. Luminescence equivalent dose, $\mathcal{L}$, is treated as a Eulerian quantity, whereby the transport of sediment by a river, the removal of luminescence by sunlight bleaching, and the erosion of new sediment are treated as conserved fluxes into and out of a control volume, $\Delta xwh$. $Q_L$ is the flux of luminescence-bearing material, $\mathcal{L}^*$ is the luminescence lost to bleaching, and $\mathcal{L}^\circ$ is the influx of new luminescence due to erosion of new sediment with accumulated charge.

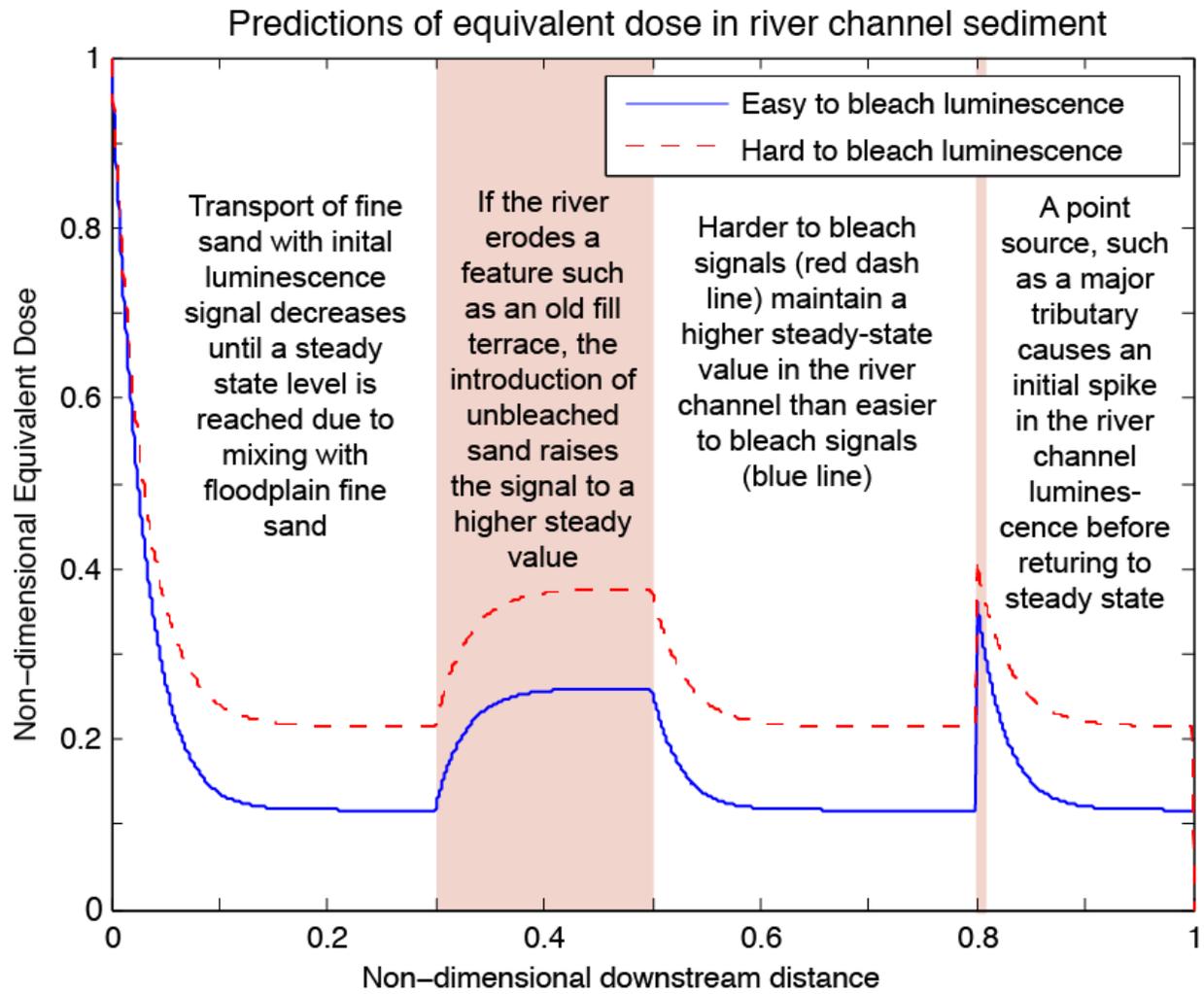

**Figure 2:** Predictions of the downstream patterns in luminescence equivalent dose of river sediment represented by Figure 1 and Equation 10. Blue line represents the model using a fast to bleach luminescence; the dashed red line shows a harder to bleach luminescence.

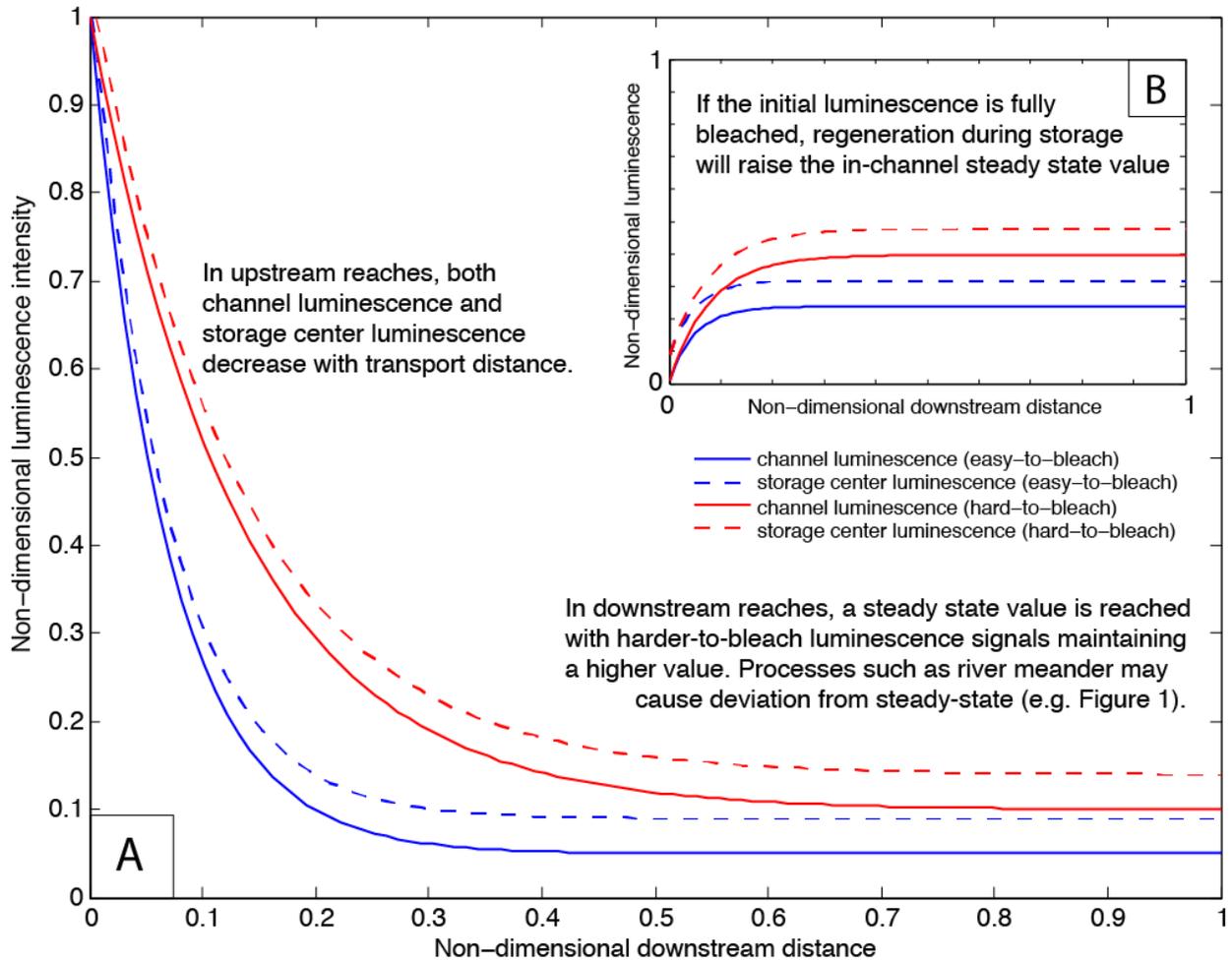

**Figure 3:** **A**) Example of a simple storage center interaction with channel sediment. Storage center is modeled with Equation 24 and transport is modeled with Equations 11 and 12. **B**) Example of a system where the initial sediment is fully bleached. Regeneration during storage causes the in-channel luminescence to increase until steady state is obtained.

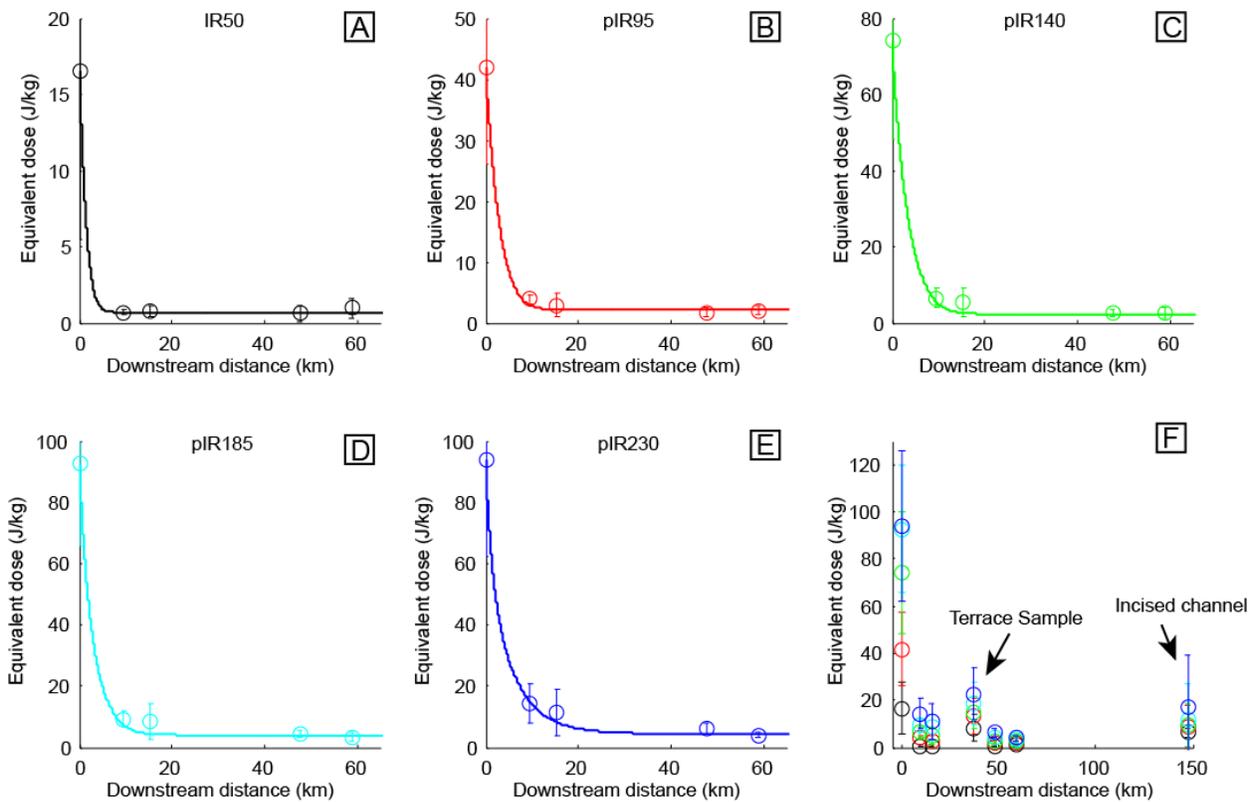

**Figure 4:** Application of the model to field data from McGuire and Rhodes [2015a] for the Mojave River in southern California, USA. **A-E)** Comparison of field data (circles with error bars) with best-fitting model run (line) for various pIR signals. **F)** Complete field data from McGuire and Rhodes data including terrace sample used for storage center luminescence and incised channel sample not included in study due to strong change in river geomorphology.

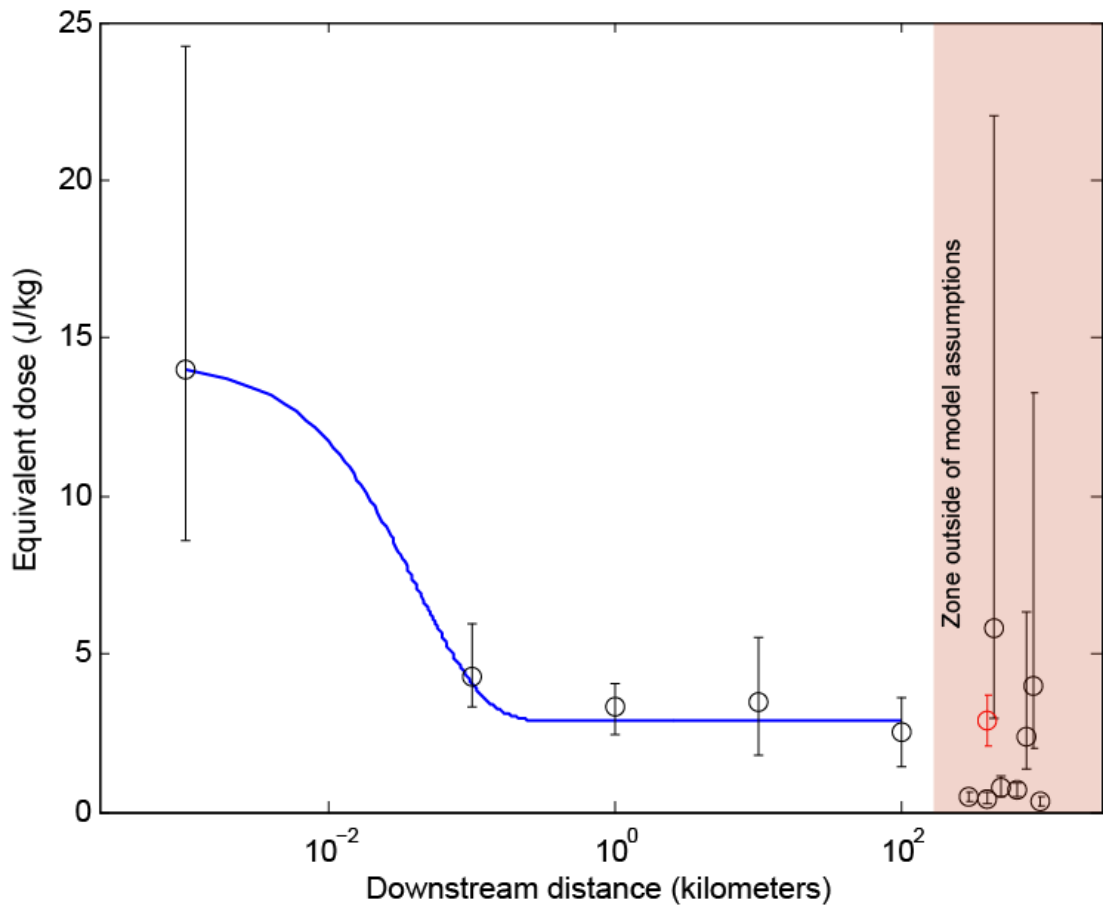

**Figure 5:** Application of the model to quartz OSL field data [Stokes et al., 2001] for the Loire in France Note logarithmic scale used for x-axis following original presentation of data in Stokes et al. [2001] and to better illustrate model/data comparison. Apparent convexity in the curve is a result of logarithmic x-axis scale. Errors on data points taken as half of the max and min reported value from Stokes et al. [2001]. Farthest downstream samples are not modeled due to uncertainties in sediment flux, storage times, and consistency of exchange rates. See text for discussion. Red circle indicates terrace data from Colls et al. [2001] used to calculate example storage timescales and exchange rates.

| River | Luminescence signal | sediment exchange rate | transport lengthscale | long-term storage timescale | time-averaged virtual velocity |
|---|---|---|---|---|---|
| | | η | ls (km) | τs (kyr) | U (km/yr) |
| Mojave River, USA | pIR230 | 11% ± 6% per km | 8.7 ± 4.5 | 5.2 ± 1.5 | 1.7 ± 1.0 |
| | pIR185 | 20% ± 10% per km | 10 ± 5.1 | 4.2 ± 1.3 | 2.4 ± 1.4 |
| | pIR140 | 17% ± 11% per km | 7.7 ± 4.7 | 3.6 ± 1.0 | 2.1 ± 1.4 |
| | pIR95 | 19% ± 7% per km | 4.8 ± 1.8 | 3.3 ± 1.1 | 1.5 ± 0.7 |
| | IR50 | 18% per km | 3.6 | 2.1 ± 0.9 | 1.7 |
| | *Average value* | *17% ± 12% per km* | *6.9 ± 4.2* | *3.6 ± 1.2* | *1.9 ± 1.4* |
| Loire River, France | quartz OSL | 4.7% ± 4.2% per m | 0.05 | 1.2 ± 0.75 | 0.04 |
| Ranges observed in previously published data | non-luminescence methods | 1.6% - 44% per km | 0.4 - 125 | 0.125 - 1.8 | 8-200 |

**Table 1:** Values obtained from application of the model to the field data of McGuire and Rhodes [2015a] for the Mojave River and the field data of Stokes et al. [2001] and Colls et al. [2001] for the Loire. Uncertainties are reported to 1-σ. Values with no uncertainty have relative errors greater than 1 and are shown for illustrative purposes.

| Symbol | Units | Description |
|---|---|---|
| *D\** | J/kg | dose growth parameter |

| $D_E$ | J/kg | equivalent dose |
|---|---|---|
| $D_R$ | J/kg/yr | background environmental dose rate |
| $f$ | s$^{-1}$ | loss rate of integrated luminescence intensity |
| $f_D$ | s$^{-1}$ | fraction of suspended sediment flux deposited per second during transport |
| $f_E$ | s$^{-1}$ | fraction of suspended sediment flux entrained from storage per second during transport |
| $f_o$ | s$^{-1}$ | unattenuated bleaching rate of integrated luminescence intensity |
| $h$ | meters | height of control volume |
| $I_0$ | arbitrary units (luminescence) | initial sensitivity-corrected integrated luminescence intensity |
| $I_{sat}$ | arbitrary units (luminescence) | sensitivity-corrected integrated luminescence intensity at saturation |
| $k_t$ | s$^{-1}$ | loss rate of equivalent dose with sunlight exposure |
| $\mathcal{L}$ | J/kg | mean equivalent dose of river sediment |

| Symbol | Units | Description |
| --- | --- | --- |
| $\mathcal{L}^*$ | J/kg/s | loss rate of equivalent dose due to sunlight exposure |
| $\ell_s$ | m | characteristic lengthscale in channel where most traveling particles have been deposited in long-term storage |
| $\ell_{system}$ | m | lengthscale of a system of interest |
| $\mathcal{L}_b$ | J/kg | mean equivalent dose of sediment in storage centers accessible by the channel. |
| $N_e$ | J | Number of trapped electrons in a control volume |
| $Q$ | J/kg | flux term describing movement of sediment with equivalent dose L |
| $q_s$ | m$^3$/s | sediment flux |
| $t_s$ | yr | time fine sand spends in long-term storage |
| $u$ | m/s | transport velocity of suspended sediment |
| $U$ | m/yr | time-averaged virtual velocity |
| $w$ | m | width of control volume |
| $x$ | m | downstream distance |
| $z^*$ | m | light attenuation coefficient representing water turbidity |
| $z_{eff}$ | m | effective depth in which a stationary grain will receive equal amounts of sunlight as a grain in turbulence (see supplemental material) |
| $\beta$ | non-dimensional | decay order of luminescence signal of interest |
| $\gamma$ | cm$^2$/photons | variable relating equivalent dose or luminescence intensity versus |
| $\eta$ | s$^{-1}$ | channel sediment / long term storage center exchange rate |
| $\lambda$ | nm | wavelength of a single color of light |
| $\tau_s$ | yr | characteristic timescale of long-term sediment storage |
| $\tau_{system}$ | yr | timescale of a system of interest |
| $\varphi$ | photons/cm$^2$/nm/s | photon flux for a given wavelength per second per cross-sectional area |
| $\varphi_0$ | photons/cm$^2$/nm/s | photon flux prior to attenuation by water |

**Table 2:** Variable units and descriptions.

ACKNOWLEDGEMENTS

We would like to thank Nathan Brown, Daniel Hobley, Amanda Keen-Zebert, Rachel Glade, and Charlie Shobe for constructive discussions. Thank you to Mayank Jain, Andrew Murray, Andrew Cyr, and Katherine Skalak for constructive comments on earlier versions of this manuscript. Thank you to editors John Buffington and Mikael Attal and three anonymous reviewers for comments with greatly improved the manuscript. HJG was supported by a Center for Integrative Research in Environmental Sciences (CIRES) fellowship. Any use of trade, product, or firm names is for descriptive purposes only and does not imply endorsement by the U.S. Government. The data used in this paper are referenced in the text. We are grateful for support from the National Science Foundation (EAR-1246546).

Harrison J. Gray[1,2]*, Gregory E. Tucker[2], Shannon A. Mahan[1], Chris McGuire[3], Edward J. Rhodes[3,4]

[1] Cooperative Institute for Research in Environmental Sciences (CIRES) and Department of Geological Sciences, University of Colorado – Boulder, CO [2] U.S. Geological Survey Luminescence Geochronology Laboratory, Denver Federal Center, Denver, CO [3] Department of Earth, Planetary, and Space Sciences, University of California Los Angeles, Los Angeles, CA [4] Department of Geography, University of Sheffield, Sheffield, S10 2TN, United Kingdom

**Contents of this file**

**Introduction**

This supplemental material file describes additional technical details used in the methods of this paper. This includes an estimation of the bleaching rate of quartz from the Loire River, a demonstration of the adequate fit of equations 14 and 15 used in this paper to describe bleaching rates, and how we estimated the uncertainty in our best-fit parameters.

**Estimation of $k_t$ for the Loire River dataset**
We follow Sohbati et al. (2011) in approximating the parameter $\varphi_o$ as the spectrum resulting from using Planck's Law for blackbody radiation. We use Planck's law in photon surface irradiance form:

$$\varphi_{sun}(\lambda) = \frac{2\pi c}{\lambda^4} \frac{1}{e^{-hc/\lambda kT}-1} \tag{S1}$$

in which λ is the wavelength of light in meters, *c* is the speed of light in meters per second, *k* is Boltzman's constant, and *T* is the temperature of the blackbody. We correct this blackbody spectrum for the spectral attenuation of light in the atmosphere (Hottel, 1976):

$$\varphi_{atmo}(\lambda) = \varphi_{sun}(\lambda) e^{-\left(\frac{\tau}{1/\cos\theta}\right)} \tag{S2}$$

$$\tau = k_a(\lambda)\, w_{H20}\, \frac{\rho_{air}}{\rho_{water}} H e^{-Z_{asl}/H} \tag{S3}$$

where θ is the solar zenith angle, τ is the optical thickness of the atmosphere and μ is an attenuation coefficient for water vapor. To account for the effects of sunlight attenuation due to weather, we normalize the solar irradiance at the top of Earth's atmosphere by the mean yearly irradiance observed across France:

$$\varphi_o(\lambda) = \varphi_{atmo}(\lambda)\left(\frac{I_{France}}{I_{Solar}}\right) \tag{S4}$$

where $I_{France}$ is the mean annual irradiance averaged from publicly available NASA global insolation data in W/m² (https://eosweb.larc.nasa.gov/cgi-bin/sse/grid.cgi), and $I_{Solar}$ is the solar insolation constant of 1368 W/m². This method allows the magnitude of photon flux to be adjusted without changing the form of atmosphere attenuated sunlight. To account for day/night cycles, we take the mean cloudless diurnal insolation (Bonan, 2002) and divide by the maximum diurnal insolation to obtain a day/night factor which we multiply by the solar photon flux.

We approximate σ(λ) by using the data of Singarayer and Bailey (2003) fitted to a function given by Bailey et al. (2011):

$$\log_{10}\sigma(\lambda) = a + b\lambda + c\lambda^2 + d\lambda^3 \tag{S5}$$

where *a*, *b*, and *c* are fitting variables. We assume that the equivalent doses measured by Stokes et al. (2001) are dominated by the fast component and therefore that the bleaching can be described using σ values for the fast component of quartz OSL (Jain et al., 2003) This simplifying assumption ignores the medium and slow components of quartz OSL, which could potentially lead to a slight overestimation of the actual bleaching rate and a slight overestimation of the sediment velocity. However, it is likely that this uncertainty is lost within the overall uncertainties in the field measurements. We evaluate the resulting equation using numerical integration. This provides the bleaching rate at the surface of the river.

Below the surface of the river, light is gradually attenuated by water and suspended sediment (Verduin et al., 1976). To account for this, we apply the Beer-Lambert law in photon-flux form for attenuation of light by water:

$$\varphi(z) = \varphi_0 e^{-k_z z(t)} \tag{S6}$$

where φ is sunlight intensity at a given depth in units of photons per second per square centimeter, *z* is depth, $\varphi_0$ is the unattenuated intensity at the water surface, and $k_z$ is an attenuation pre-exponential factor that depends on suspended sediment concentration and represents the turbidity of the river (Verduin et al., 1976). Estimating this parameter is challenging because river turbidity is often only determined empirically and no turbidity data are available for the Loire River. We approach this issue by combining the concentrations predicted by the Rouse sediment concentration profile (Rouse, 1937) with the derivation for the attenuation coefficient $k_z$. This allows for the parameter to be estimated based on assumptions of the river depth and basal sediment distributions rather than empirical extrapolations. The Beer-Lambert law is developed from

$$\partial \varphi(z) = -\mu(z)\varphi(z)\partial z \tag{S7}$$

where $\varphi$ is the photon flux, $z$ is depth, and $\mu$ is the attenuation coefficient for an individual absorbing species in units of inverse meters. Here $\partial z$ is assumed to be small enough that one particle cannot obscure another particle within the $\partial z$ layer. The attenuation coefficient, $\mu$, is simply the absorbing cross-sectional area of a particle times the number of particles within a volume defined by $\partial x \partial y \partial z$:

$$\mu_p(z) = \frac{A_p N_p}{\partial x \, \partial y \, \partial z} \tag{S8}$$

$$k_z = \sum_{i=1}^{\infty} \mu_i(z) \tag{S9}$$

where $A_p$ is the cross-sectional area of a particle and $N_p$ is the number of particles. $K_z$ is defined as the summation of all particles across suspended grain sizes with each distinct grain size indicated by the subscript, $i$. We assume spherical particles and that the number of particles can be represented by the total volume of particles divided by the volume of a particle:

$$N_p = \frac{C_i \partial x \partial y \partial z}{\frac{4}{3}\pi \left(d_i/2\right)^3} \tag{S10}$$

where $C_i$ is the concentration of particle $i$ with diameter $d_i$. The concentration $C_i$ is given by the Rouse sediment concentration profile [Rouse 1937]:

$$C_i(z) = C_i(z_0)\left[\frac{z}{h-z}\frac{h-a}{a}\right]^{R_i} \tag{S11}$$

$$R_i = \frac{\omega_{fi}}{\beta_i k u_*} \tag{S12}$$

where $h$ is the river depth, $C_i(z_0)$ is the concentration at a reference height $z_0$, R is the Rouse Number described by $W_f$, the particle settling velocity, u* the shear velocity, κ, von Karmen's constant, and β, the ratio of ratio of fluid to particle momentum transfer. This allows one to solve for the attenuation coefficient $k_z$ based on assumptions of river depth and the suspended particle size distribution at the bed described by $C_i$. We limit the grain sizes to those with Rouse numbers in the range of 0-0.5 as these will be dominantly suspended grains that will attenuate light in the water column. We use a basal concentration formula to calculate $C_i(a)$ (McLean, 1992) and estimate the distribution of grain sizes from bed load analysis on the Loire River (Gautier and Peters, 2002). We do not include the effects of scattering or refraction on the light attenuation with depth (Sassaroli and Fantini, 2004) and we further assume that additional turbidity due to dissolved organic matter is negligible. These effects would cause greater light attenuation and lower the bleaching rate. Direct measurements of the turbidity and light attenuation of the Loire River would improve the accuracy of $K_t$ estimation.

For sedimentary grains in suspension, grain paths will be highly stochastic owing to the nature of fluid turbulence. The depth at any given time will vary randomly. However, grain paths for suspended sediment are stochastic enough such that grains should spend similar amounts of time at similar depths for a single grain size. In the case of sediment used for luminescence dating (63 -250 μm), these grains have Rouse numbers on the order of ~0.25-0.5, indicating dominantly suspended transport. These grains will spend some time at the surface where they bleach rapidly, and some time at depth where minimal bleaching occurs.

For these highly stochastic paths, z(t) can be represented by an effective z, defined such that a grain traveling exclusively at this level would produce the same average luminescence change as a large ensemble of grains undergoing random vertical motions due to turbulence. We determine this effective z by setting the total light exposure for the stochastic z(t) equal to the effective z:

$$Light_{total} = Light_{effective} \tag{S13}$$

The total light exposure can be broken down into time multiplied by the light intensity at the depths at which the grain is traveling. The $Light_{effective}$ term is simply time multiplied by the light intensity described by the Beer-Lambert law at the effective z. The stochastic light intensity is described by the integral of the Beer-Lambert law with a time-varying z(t) times the probability distribution of finding a grain at a given z (p(z)).

$$t \int_0^h p(z) \varphi_0 e^{-k_z Z(t)} \partial z = t \varphi_0 e^{-k_z Z_{effective}} \tag{S14}$$

The probability distribution for a grain in suspension can be described as the sediment concentration normalized by the maximum concentration near the base of the riverbed. This can in turn be described using a Rouse sediment concentration profile for the grain size of interest [Rouse 1937]:

$$p(z) = \frac{C(z)}{C(z_0)} = \left[\frac{z}{h-z} \frac{h-z_0}{z_0}\right]^R \tag{S15}$$

$$R = \frac{\omega_f}{\beta k u_*} \tag{S16}$$

where C(z) is the concentration at depth z, C(a) is the reference concentration at a height a near the bed, h is the height of the flow, R is the Rouse number, $W_f$ is the sediment fall velocity, β is the ratio of fluid to sediment momentum transfer, κ is von Karmen's constant, and $u_*$ is the river shear velocity.

Solving for $z_{effective}$ results in the following equation:

$$Z_{effective} = \left[\frac{1}{-k_z}\right] ln\left[\int_{z_0}^h \left[\frac{z}{h-z} \frac{h-z_0}{z_0}\right]^R e^{-k_z Z(t)} \partial z\right] \tag{S17}$$

which must be solved via numerical integration. This $z_{effective}$ can then be used to estimate the $k_t$ and solve for the bleaching term L*.

**Bleaching experiment to test Equations 14 and 15**

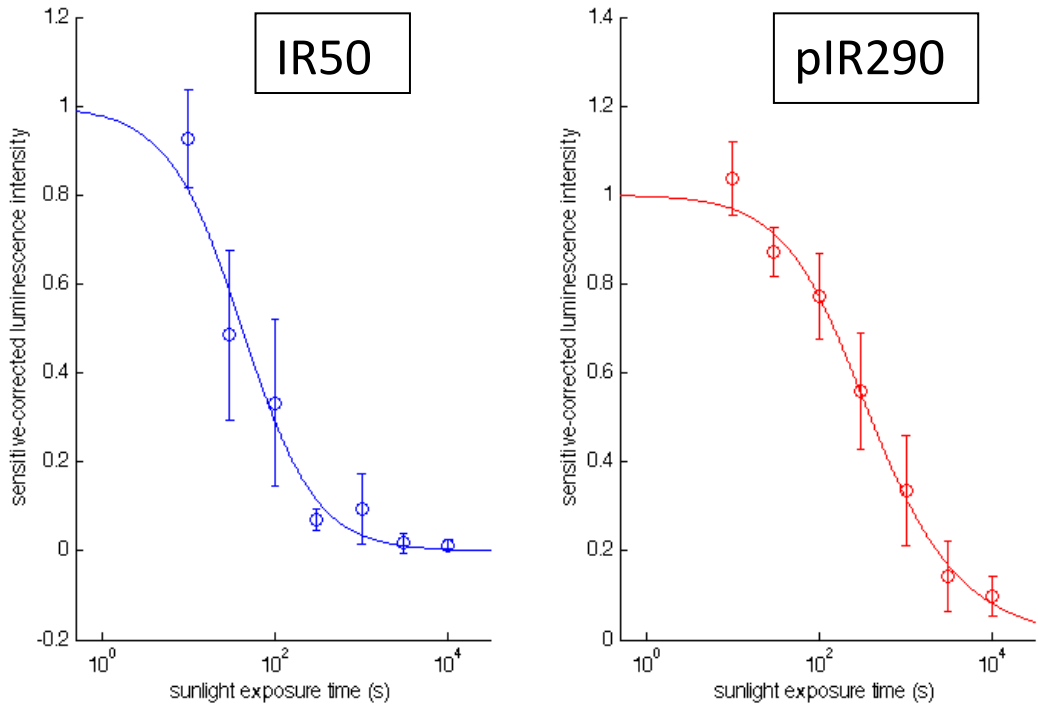

**Figure S1:** Figure showing fit of equations 14 and 15 to sunlight bleaching experiments. See supplemental text for details.

To test if Equations 14 and 15 adequately describe bleaching, we performed simple bleaching experiments on large aliquots of 90-250 µm feldspar sand from the South River near Waynesboro, VA (Figure S1). Left graph shows IR50; right graph shows pIR290; circles and error bars show mean of 6 aliquots, lines show best-fit application of Equations 14 and 15. Aliquots were dosed with 70 Gray and separately exposed to natural sunlight for 10, 30, 100, 300, 1000, 3000, and 10,000 seconds. We used 6 aliquots per time increment. The equations provide an adequate fit to the data, whether the signal is fast to bleach (pIR50) or slow to bleach (pIR290) and we interpret that this equation can provide an empirical approximate of general luminescence bleaching.

**Uncertainty Analysis**

To determine the controls and uncertainty in *u* and *L°*, we apply the general formula for error propagation (Ku et al., 1966):

$$Q = f(x, y, z) \tag{S18}$$

$$\sigma_Q^2 = \sigma_x^2 \left(\frac{\partial Q}{\partial x}\right)^2 + \sigma_y^2 \left(\frac{\partial Q}{\partial y}\right)^2 + \sigma_z^2 \left(\frac{\partial Q}{\partial z}\right)^2 \tag{S19}$$

for a given function *Q* that depends on variables *x, y, z*. The uncertainty in *Q* described by $\sigma_Q$ is calculated by the summation of the squares of the uncertainties in each variable multiplied by the partial derivatives of Q with respect to each variable. We solve these equations for the uncertainty in u and η.

We also evaluate the uncertainty in the least-squares fit by using the equation (Bevington and Robinson, 2003):

$$\sigma_\mathcal{L}^2 = \frac{1}{N-n}\Sigma_{i=1}^{N}(\mathcal{L}_i - \mathcal{L}_m(u,\eta))^2 \tag{S20}$$

where $\mathcal{L}_i$ is the field data and $\mathcal{L}_m$ is the modeled data, N is the number of field data points, and n is the number of model parameters.

SUPPLEMENTAL INFORMATION REFERENCES